# ROTATIONAL SPECTRA OF THE BARYONS AND MESONS

David Akers*
*Lockheed Martin Corporation, Dept. 6F2P, Bldg. 660, Mail Zone 6620,*
1011 Lockheed Way, Palmdale, CA 93599
*Email address: David.Akers@lmco.com

**Abstract**

An investigation of the rotational spectra of baryons and mesons is conducted. Diakonov, Petrov and Polyakov claimed that all light baryons are rotational excitations. A study of the history of particle physics indicates that the ideas of rotational spectra can be originally attributed to a constituent-quark (CQ) model as proposed by Mac Gregor. Later research advanced spin-orbit splitting in a deformed model as suggested by Bhaduri and others. In the present work, we show from current data that the rotational spectra of baryons and mesons are in agreement with the original claims of Mac Gregor: namely, the values for the rotational energies $E_{rot}$ of particles merge with those of nuclear rotational bands in light nuclei. It is also shown that particles of different isotopic spins are separated in mass by a 70 MeV quantum, which is related to the SU(3) decuplet mass spacing as originally proposed by Gell-Mann.

## INTRODUCTION

In a study of exotic baryons, Diakonov, Petrov and Polyakov claimed that all light baryons are rotational excitations [1]. In the Skyrme model, nucleons are considered to be solitons of a pion field, and the excitation energies for the rotational states are:

$$E_{rot}(J) = (\hbar)^2 J(J+1)/2I, \qquad (1)$$

where $I$ is the moment of inertia. The soliton idea is only one of several models, which propose rotational states for the baryons and mesons. A non-relativistic quark model has been proposed by Bhaduri *et al.*, which has similar ideas of rotational states and which has valence quarks moving in a deformed oscillator potential [2]. These ideas were later expanded for strange baryons in a deformed baryon model [3]. Hosaka and others have also shown with success that their deformed oscillator quark model fits baryon rotational states [4]. However, these models are not the earliest theories for the proposal of rotational states in particle physics. In fact, the earliest known research into a study of the rotational spectra of baryons and mesons can be attributed to Mac Gregor from the 1970s [5-9]. Mac Gregor also studied particle deformations.

In the work of Mac Gregor, a constituent-quark model was proposed that featured baryons and mesons as rotational states, which suggested a correlation between increasing L values and increasing energies in these resonances. These resonances followed the expected $L(L + 1)$ interval rule for rotational spectra, and the experimental values for the rotational energies of hadrons merged into the $A^{-5/3}$ scaling law for light nuclei with compact or prolate geometries. In the present work, we present evidence with data from the most recent *Review of Particle Properties* [10] and suggest that the original claims by Mac Gregor are correct. The main thesis of this paper is that excited baryons and mesons are rotational states sitting on a quark substructure, which also has a 70 MeV quantum for energy separations. We show that baryons and mesons are in fact separated in multiples of 70 MeV masses in agreement with the earlier proposals of Mac Gregor. It is suggested that the 70 MeV quantum is related to the SU(3) decuplet spacing as originally proposed by Gell-Mann in the quark model.

## HADRONIC ROTATIONAL BANDS

In the early 1970s, Mac Gregor demonstrated how to accurately bridge the gap between hadronic and nuclear domains. By an application of nonadiabatic rotational bands, Mac Gregor determined a proper understanding of particle resonances which is applied today by others [2-3]. The nature of the nonadiabatic rotations starts from the rotational Hamiltonian [8]:

$$H = (\hbar)^2 \mathbf{L}^2/2I, \qquad (2)$$

where $\mathbf{L}$ is the angular momentum operator and $I$ is the moment of inertia. The angular momentum operator can be written as

$$\mathbf{L} = \mathbf{J} - \mathbf{S}, \qquad (3)$$

where $\mathbf{J}$ and $\mathbf{S}$ are the total angular momentum and spin angular momentum operators. The angular momentum operator in Eq. (2) can be written as

$$\mathbf{L}^2 = (\mathbf{J} - \mathbf{S})^2 = \mathbf{J}^2 + \mathbf{S}^2 - 2\mathbf{J} \cdot \mathbf{S}. \qquad (4)$$

Mac Gregor noted that the Coriolis term $\mathbf{J} \cdot \mathbf{S}$ tended to vanish if the rotating system satisfied a few conditions [8]. One of these conditions is the intrinsic spin S does not have the value ½. Mesons would fit into this category (see Section 10 in Ref. 11). If the Coriolis term vanishes, then the rotational energies of the mesons would follow a J(J + 1) interval rule:

$$E(J) = E_0 + (\hbar)^2 J(J+1)/2I,$$

$$E_{rot} = (\hbar)^2/2I, \qquad (5)$$

where $E_0$ is the bandhead energy. The bandhead energy is the ground state of the particular meson series.

On the other hand, when the Coriolis term does not vanish for rotational levels, the particle resonances follow the L(L + 1) interval rule and the rotational energies are

$$E(J) = E_0 + (\hbar)^2 L(L+1)/2I,$$

$$E_{rot} = (\hbar)^2/2I. \qquad (6)$$

The baryons therefore follow the L(L + 1) interval rule of Eq. (6). Likewise, Mac Gregor showed Eq. (6) is the correct equation for all light nuclei with A ≤ 20 [8].

Before we start plotting the rotational spectra of baryons and mesons, there are a few preliminary topics for discussion. First, how do we sort the resonances to identify a particular rotational series? Second, what is the energy scale for sorting these resonances? Third, how does this energy scale fit into the current Standard Model of particle physics? The first question is answered by sorting on angular momentum L values for baryons and total angular momentum J values for mesons. The second is answered by a study of the experimental systematics [7, 12-13]. In Ref. 12 and references therein, there are clear indications of a 70 MeV quantum in the excitations of baryons and mesons. The energy scale of 70 MeV has been utilized to build *Isotopic Tables of the Baryons and Mesons* [13]. The energy scale is labeled with notation m = 70 MeV, B = 140 MeV, F = 210 MeV and X = 420 MeV in the constituent-quark (CQ) model of Mac Gregor [12]. These are the labels that we choose for the tabulation of all baryon and meson resonances in the present work.

In a study of exotic baryons [14], it is shown that there are well-known mass relations in the Standard Model. The SU(3) decuplet has an equal spacing rule [14]:

$$M_\Omega - M_{\Xi^*} = M_{\Xi^*} - M_{\Sigma^*} = M_{\Sigma^*} - M_\Delta ,  \qquad (7)$$

where this spacing is equal to approximately 140 MeV. It is obvious that $M_\Omega - M_{\Xi^*} = B = 2m$. Thus, the SU(3) decuplet is placed in the *same* row in the isotopic table of the baryons [13]. In order to account for new exotic baryons, new columns are opened in the isotopic table of the baryons [13]. These columns correspond to the equal spacing rule for the baryon anti-decuplet [14]. Additional columns N + 5m or N + 7m may be opened in the isotopic table of the baryons for exotic I = 3/2 cascade baryons, where N = 939 MeV for the nucleon, and m = 70 MeV. This may suggest that the anti-decuplet baryons are shifted from the SU(3) decuplet by the energy scale of m = 70 MeV.

We are now in a position to tabulate all the known and not-so-well established baryon resonances from a sort on angular momentum L and on an energy scale with a 70 MeV grid. In Table I, we list the measured values of nucleon masses as a function of angular momentum L. The CQ Model notation is in the left most column of Table I. The calculated bandhead energies $E_{calc}$ are based upon the constituent-quark model of Mac Gregor. Rotational energies are calculated between adjacent states and are averaged under the column for $E_{rot}$. Predicted states are indicated in red color. An example of a predicted state, the N(1359)$S_{11}$, has supporting experimental evidence [15]. In Table I, the N(1004) state has some recent preliminary evidence for its existence [16]. The nucleon resonances of Table I are plotted in Fig. 1. In Fig. 1, the black diamonds represent N(939) and N(1004). The dash line between these two states indicates a preliminary status. The data with circles represent the nucleons N(1359)$S_{11}$, N(1440)$P_{11}$, N(1520)$D_{13}$, N(1680)$F_{15}$, N(1910)$G_{17}$, and N(2220)$H_{19}$. A simple calculation will show that the experimental resonances are within 2-3% of the theoretical values postulated by Eq. (6). In Fig. 1, the squares represent the row starting with N(1534)$S_{11}$ in Table I. The triangles represent the row starting with N(1650)$S_{11}$, and the asterisks represent the row starting with N(2090)$S_{11}$. The average rotational energy of the nucleons is calculated to be 28.5 MeV.

In Table II, the Δ baryons are tabulated with the same notation as in Table I. The rotational spectra of the Δ baryons are plotted in Fig. 2. The squares represent the row starting with Δ(1540)S$_{11}$ in Table II. The triangles represent the row starting with

Δ(1620)S$_{11}$, and the crosses represent those baryons in the row starting with Δ(1690)S$_{11}$.

The average rotational energy of the delta baryons is calculated to be 31.8 MeV.

In Table III, the Λ baryons are tabulated with the same notation as in Table I. The

rotational spectra of the Λ baryons are plotted in Fig. 3. The squares represent row

starting with Λ(1326)S$_{11}$, the triangles represent the row starting with Λ(1405)S$_{11}$, and the

crosses are associated with the row starting at Λ(1466). The diamond symbols represent

the row starting with Λ(1670)S$_{11}$, and the circles represents the row starting with

Λ(1800)S$_{11}$. The average rotational energy of the lambda baryons is calculated to be 30.4

MeV.

In Table IV, the Σ baryons are tabulated again with the same notation. The rotational

spectra of the Σ baryons are plotted in Fig. 4. The circles represent the row starting with

Σ(1480)S$_{11}$, the squares are the data from the row starting with Σ(1541)S$_{11}$, the triangles represent those sigma baryons in the row starting at Σ(1620)S$_{11}$, and the diamonds represent those from the row starting at Σ(1750)S$_{11}$. The average rotational energy of the sigma baryons is calculated to be 30.1 MeV.

In Table V, the Ξ baryons are listed with the same notation as in Table I. The amount of available data is limited to one row starting with Ξ(1671)S$_{11}$. This single row of three data points is plotted in Fig. 5. For the Ω baryons, the amount of data is very limited.

The rotational energies for the Ω baryons are calculated in Table VI.

From Tables I through VI, we plot the calculated values of E$_{rot}$ for the baryons as a function of the bandhead mass. These results are shown in Fig. 6. In Fig. 6, the horizontal black line represents the average rotational energy of all baryons. This is calculated to be 30 MeV. The blue diamonds in Fig. 6 represent the calculated values for the nucleons. We note that the nucleons are sloping downward with increasing bandhead mass. This is in agreement with the earlier findings of Mac Gregor [8]. The red circles are the Δ baryons in this figure. The triangles are the Λ baryons. The crosses represent the Σ baryons, the asterisks are the Ξ baryons, and the circles represent the Ω baryons. Our Fig. 6 shows the most current data based upon those taken from the *Review of Particle Properties (2002)* [10]. This figure is consistent with Mac Gregor's work (see Fig. 17 in Ref. 8).

We now turn to a study of the rotational spectra of mesons. The isospin I = 0 mesons are tabulated in Table VII. The difference between the baryons and the mesons is the fact that mesons are sorted with total angular momentum J compared to L values for the baryons. The rotational energies $E_{rot}$ of mesons are based upon Eq. (5), and the spectra follow the J(J + 1) rule. In the earlier work of Mac Gregor [9], the ω(782) and ϕ(1020) are attributed to be spinors with J = 1. These mesons are position holders in Table VII. Earlier research indicated the possible existence of $A_1^L$(1050) as noted in Fig. 2 of Ref. 17. A tremendous amount of low-mass mesons, studied by Mac Gregor, is no longer listed in the *Review of Particle Properties* (RPP). We therefore use only the currently available data from the RPP to establish rotational properties for mesons and to draw conclusions.

In Fig. 7, the rotational spectra of the mesons (I = 0) are plotted. The circle with the blue line represents the row with η'(958) and ϕ(1020). The squares with the green line represent the row starting with $f_0$(1370), the asterisks are those mesons starting with

η(1440), and the triangles represent the row of data starting with $f_0(1500)$. The diamond symbols are those mesons in the row starting at 1610 MeV. The circles with brown line represent the row starting with $f_0(1710)$, and the crosses are those starting in the row with a predicted η(1820) meson.

The kaons (I = ½) are tabulated in Table VIII and plotted in Fig. 8. In Fig. 8, the K(892) is shown. The open circle represents a non-existent kaon at J = 0. The line with + symbols represents the row starting at 1214 MeV in Table VIII. The squares on a green line represent the kaons starting with the level at 1354 MeV. The next row, starting with K(1430), is represented by the crosses and black line. The triangle symbols represent the row starting at 1543 MeV. Finally, the diamonds represent the kaons starting with K(1950).

For the light-mass mesons (I = 1), we have tabulated these in Table IX and plotted them in Fig. 9. The ρ(770) is a clear rotational resonance in the CQ Model of Mac Gregor and is indicated as shown in Fig. 9. The + symbols represent those mesons starting at the 1115 MeV level. The triangles represent the row starting with π(1300). The squares on the green line represent the row at the 1395 MeV level. The circles on the brown line represent the kaons starting at 1535 MeV. Finally, the diamonds on the blue line represent the row starting at 1675 MeV.

The calculated values of $E_{rot}$ for mesons are added to those for baryons from Fig. 6; these results are plotted in Fig. 10. The + symbols represent the mesons with isospin I = 0. The open squares are the kaons (I = ½), and the closed squares are the meson with I = 1. As we mentioned earlier, the low-mass mesons have disappeared over the decades from the compilation in the *Review of Particle Properties*. Therefore, we expect our results to be slightly different from those of Mac Gregor. Overall, the rotational energies of the mesons are lower than previous calculations by Mac Gregor, but the low-mass mesons are still above the average rotational energy for the baryons; they are roughly 40 MeV in rotational energy for bandhead masses around 770 MeV.

As a final exercise in this section, we plot the calculated values of $E_{rot}$ for mesons only and show the $A^{-5/3}$ curve for the nuclear domain. This is shown as Fig. 11. The blue line and squares are the isospin I = 1 mesons. The other symbols of the mesons are the same as those in Fig. 10. The dashed line is the $A^{-5/3}$ curve from Mac Gregor [8]. We show a few nuclei He-3, Li-5 and C-12 from Mac Gregor [7]. The current data of meson resonances from the *Review of Particle Properties* [10] are consistent with the merging of the hadronic realm into the nuclear domain, involving those light nuclei with compact or prolate geometries. Thus, we have verified the earlier results of Mac Gregor.

Moreover, we have also calculated the rotational energies for charmed and bottom mesons. The results are tabulated in Tables X and XI. The amount of data is very limited. The rotational energies of the charmed mesons are also plotted in Fig. 11; these are shown as orange triangles. The orange line is suggestive of a slope, which is downward pointing to the Li-5 nucleus. Likewise, we have plotted the rotational energies for the extremely limited data available for the bottom mesons. They are shown as black circles. The slope of the black line breaks downward pointing to the C-12 nucleus. The author has also calculated the rotational energies for the D mesons and for charmed $\Xi_c$ baryons. These resonances have a slope of their line downward pointing to the He-3. For clarity, we do not show these data points in Fig. 11. The extension of the CQ Model to these mesons (as in Tables X and XI) may be limited; even though these data points are suggestive of a link to the nuclear domain, we must be cautious regarding these data points in Fig. 11. Mac Gregor calculated higher rotational energies for these mesons from a relativistic spinning sphere model (see page 1020 in Ref. 12).

## ISOTOPIC TABLES OF THE BARYONS AND MESONS

The results from our analysis on the rotational spectra of baryons and mesons have been listed in the revised *Isotopic Tables of the Baryons and Mesons (2003)* [13]. From the tables, we would like to present evidence of the 70 MeV quantum as suggested by Mac Gregor. The baryon spectrum from the isotopic table is shown in Fig. 12. For each isospin, core masses (with $J = \frac{1}{2}$) are subtracted from each baryon resonance in the vertical columns of the baryon table in Ref 13. The differences are then plotted in Fig. 12. The baryons are represented by colors in Fig. 12: N(green), Δ(blue), Λ(pink), Σ(red), Ξ(purple), and Ω(yellow). The black-dashed lines represent baryons, which are predicted to exist. In Fig. 12, there is noted some evidence of precise mass quantization (m = 70 MeV) in the high angular momenum *L*- values (total $J = L + S$). We emphasize here that the mass differences in Fig. 12 are derived from the experimental data [10] and that the m = 70 MeV energy separations are real.

Likewise, we can plot the mass differences for mesons. From the *Isotopic Table of the Mesons* [13], we have calculated core corrections for the mesons. For each isospin, core masses from the ground states are subtracted from each meson resonance in the vertical column of the meson table. The differences are then plotted in Fig. 13. In Fig. 13, the green colored levels are I = 1 meson differences with respect to π (135). The magenta colored levels represent I = ½ meson differences with respect to K (494), and the blue colored levels indicate I = 0 levels with respect to η (547).

In Fig. 13, it can be easily noted that there are patterns of energy separations: F = 210 MeV, BF = 350 MeV, and X = 420 MeV. As we mentioned in the introduction, this notation comes from the CQ Model of Mac Gregor [12]. These energy differences are *experimental* values and are suggestive of a multiple of the 70 MeV quantum.

## CONCLUSION

In this study of the rotational spectra, we have shown that light baryons are indeed rotational excitations. Moreover, the rotational energies of the mesons also overlap with those of the baryons. We have shown that the earliest known research into a study of the rotational spectra of baryons and mesons can be attributed to Mac Gregor from the 1970s [5-9]. The baryon resonances followed the expected L(L + 1) interval rule for rotational spectra while the meson resonances followed the J(J + 1) rule. We have shown that the experimental values for the rotational energies of hadrons merged into the $A^{-5/3}$ scaling law for light nuclei with compact or prolate geometries. This is in agreement with the earlier findings of Mac Gregor. In the present work, we presented evidence with data from the most recent *Review of Particle Properties* [10] and suggested that the original claims by Mac Gregor are correct. The main thesis is verified that excited baryons and mesons are rotational states sitting on a quark substructure, which also has a 70 MeV quantum for energy separations. We have shown that baryons and mesons are in fact separated in multiples of 70 MeV masses in agreement with the earlier proposals of Mac

Gregor. It was suggested that the 70 MeV quantum is related to the SU(3) decuplet spacing as originally proposed by Gell-Mann in the quark model.

## ACKNOWLEDGEMENT

The author wishes to thank Dr. Malcolm Mac Gregor, retired from the University of California's Lawrence Livermore National Laboratory, for his encouragement to pursue the CQ Model, and he wishes to thank Dr. Paolo Palazzi of CERN for his interest in the work and for e-mail correspondence.

**FIGURE CAPTIONS**

Fig. 1. The rotational spectra of the nucleons are shown as a function of the $L(L + 1)$ interval rule.

Fig. 2. The rotational spectra of the delta baryons are shown as a function of the $L(L + 1)$ interval rule.

Fig. 3. The rotational spectra of the lambda baryons are shown as a function of the $L(L + 1)$ interval rule.

Fig. 4. The rotational spectra of the sigma baryons are shown as a function of the $L(L + 1)$ interval rule.

Fig. 5. The rotational spectra of the cascade baryons are shown as a function of the $L(L + 1)$ interval rule.

Fig. 6. Calculated values of the rotational energies for the baryons are shown as a function of the bandhead mass.

Fig. 7. The rotational spectra of the mesons ($I = 0$) are shown as a function of the $J(J + 1)$ interval rule.

Fig. 8. The rotational spectra of the kaons ($I = ½$) are shown as a function of the $J(J + 1)$ interval rule.

Fig. 9. The rotational spectra of the mesons ($I = 1$) are shown as a function of the $J(J + 1)$ interval rule.

Fig. 10. Calculated values of the rotational energies for the baryons and mesons are shown as a function of the bandhead mass.

Fig. 11. Calculated values of the rotational energies for only the mesons are shown as a function of the bandhead mass. Charmed and bottom mesons are indicated as orange and black data points, respectively. Light-mass mesons are the same as those in Fig. 10.

Fig. 12. The baryon spectrum, with core subtractions, is shown. The indicated data are discussed in the text. The baryons are represented by color: N(green), Δ(blue), Λ(pink), Σ(red), Ξ(purple), and Ω(yellow).

Fig. 13. The meson spectrum, with core subtractions, is shown. The indicated data are discussed in the text. I = 1 mesons are green color. I = ½ kaons are magenta color, and I = 0 meons are blue color.

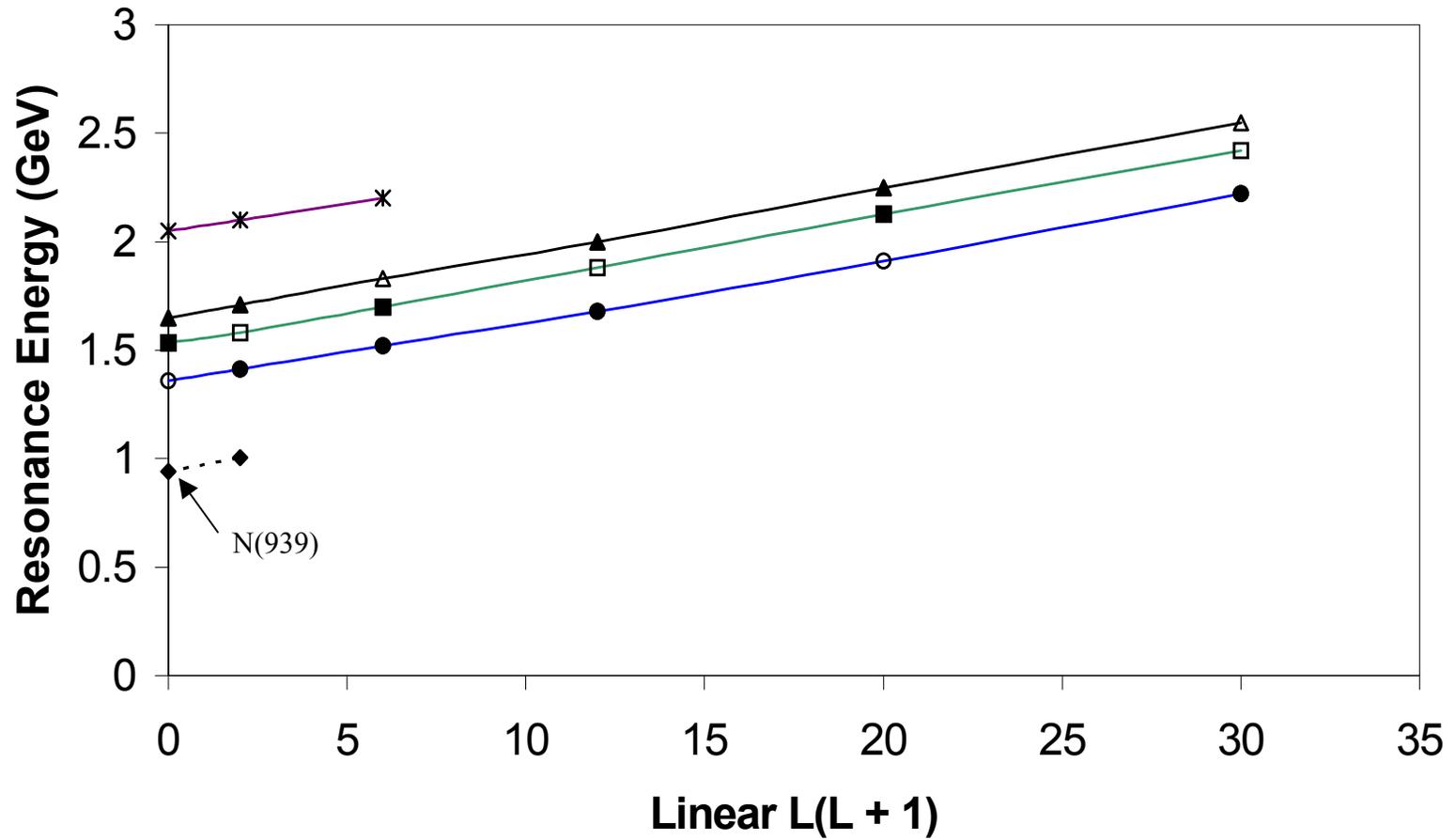

Fig. 1

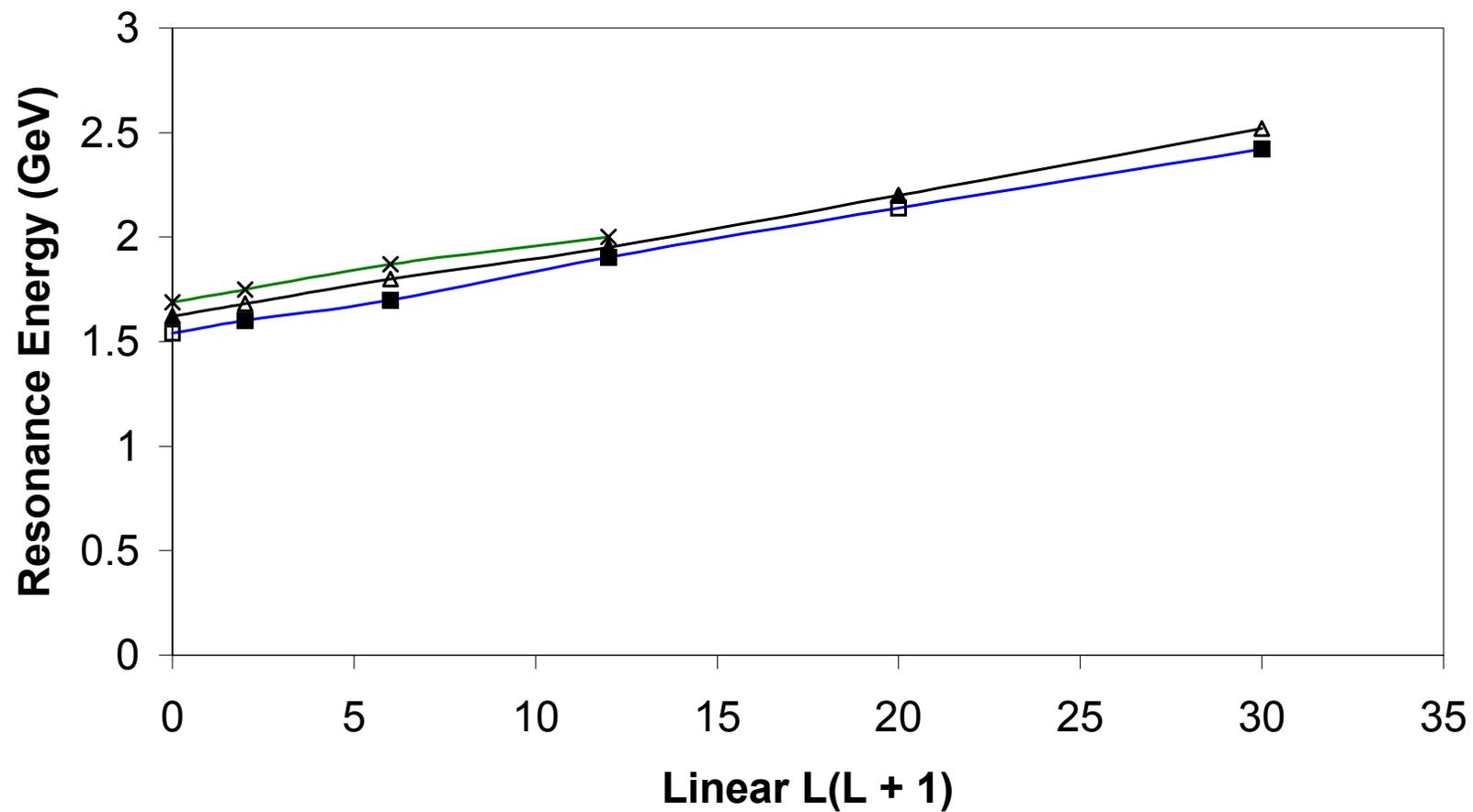

Fig. 2

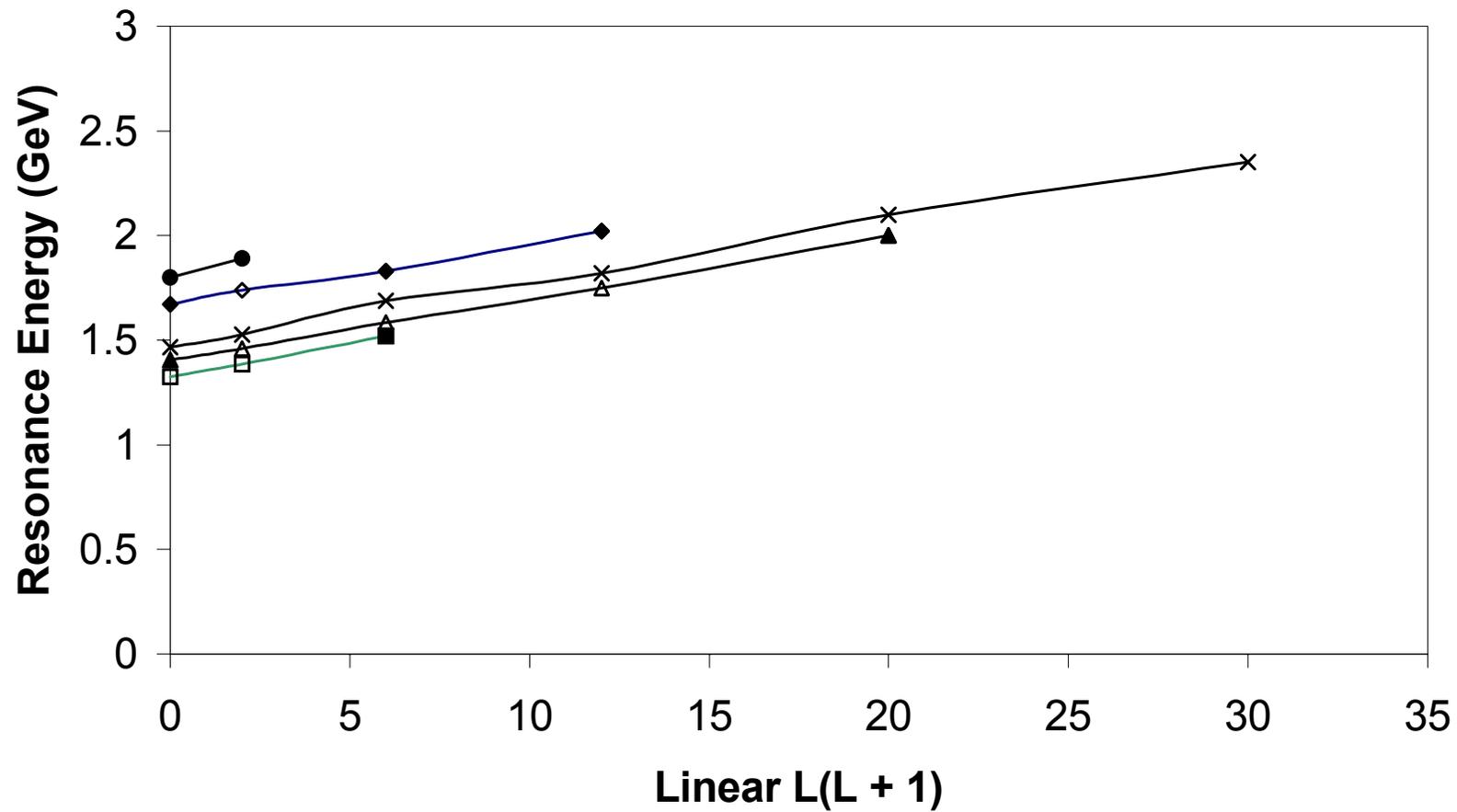

Fig. 3

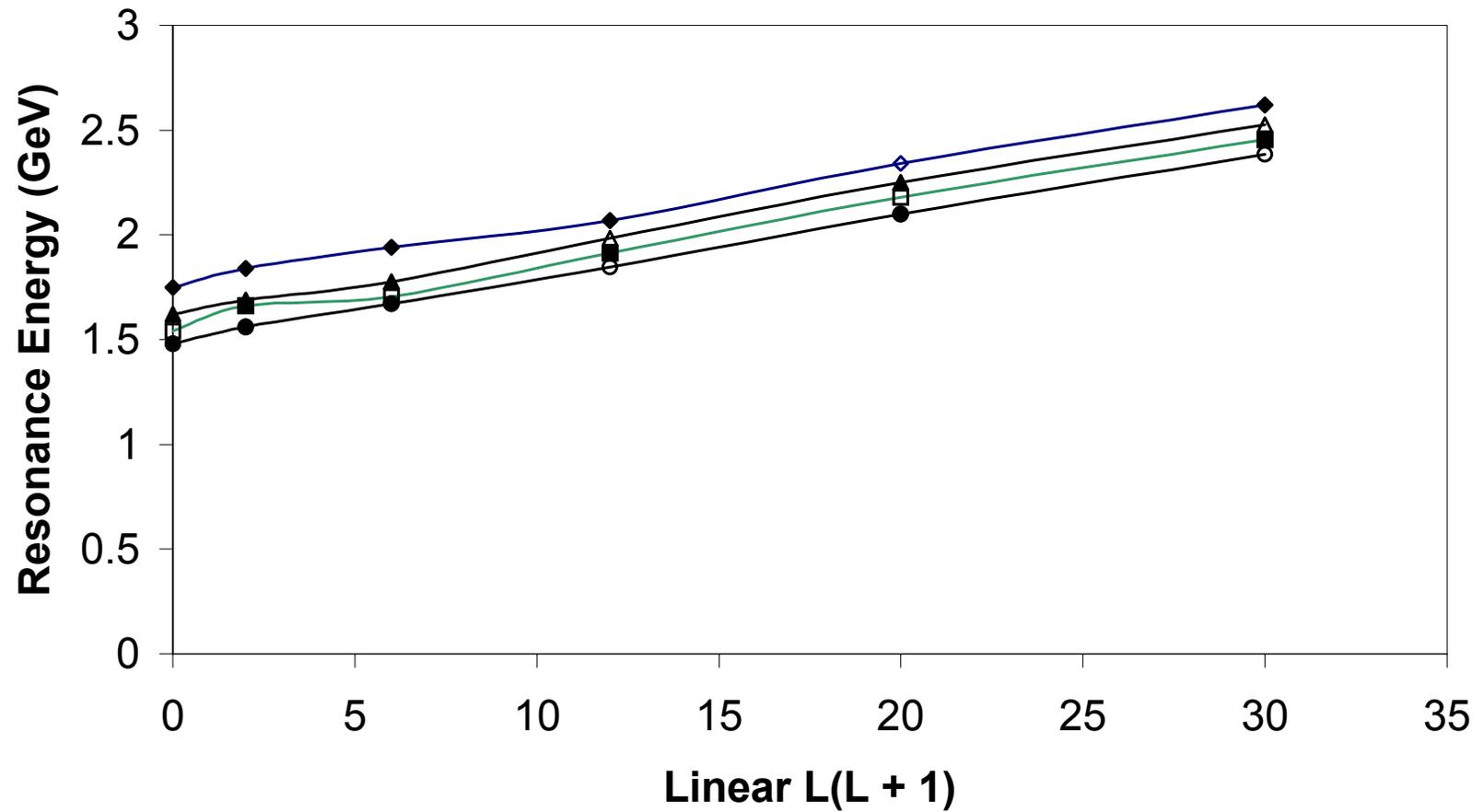

Fig. 4

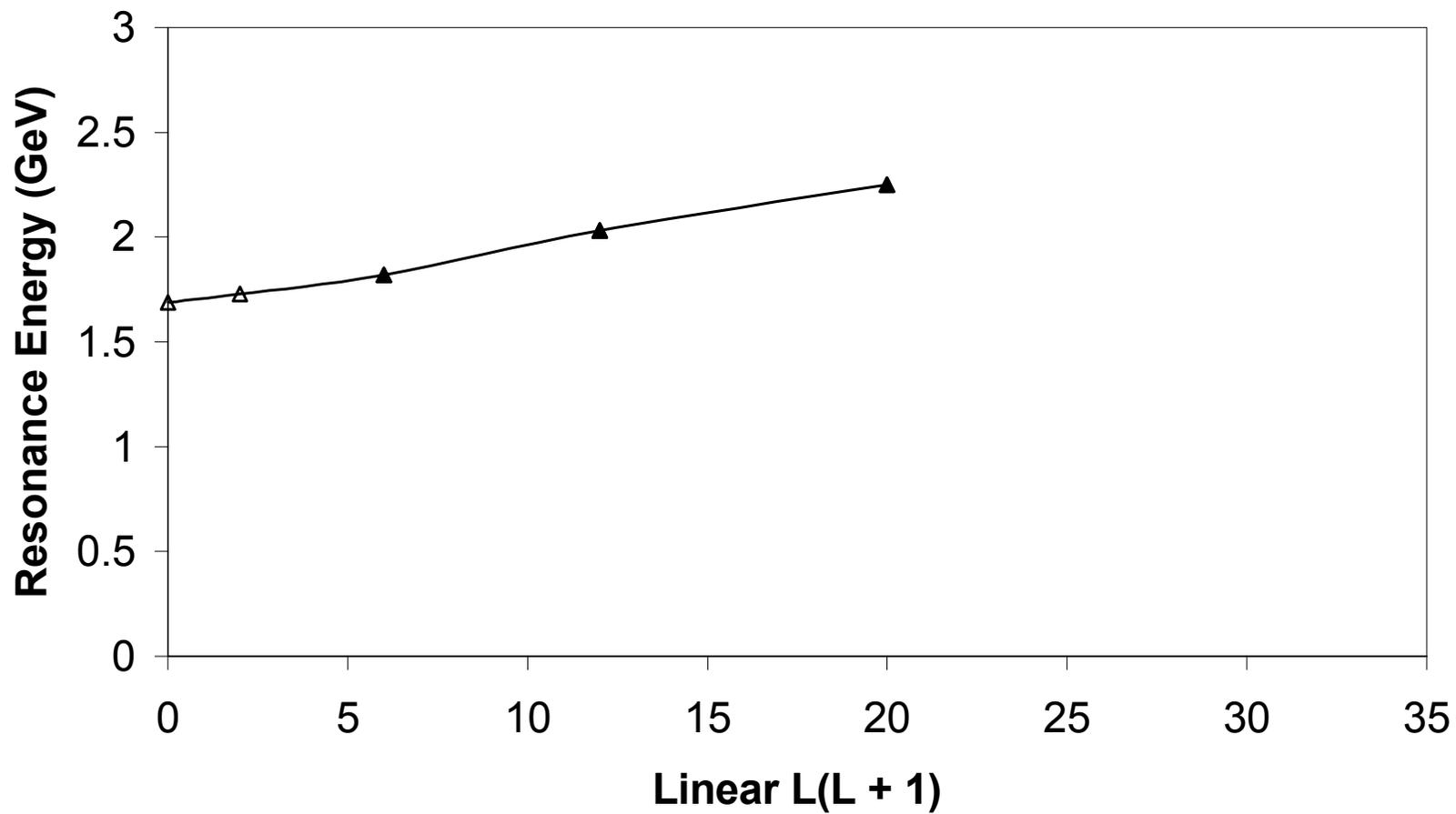

Fig. 5

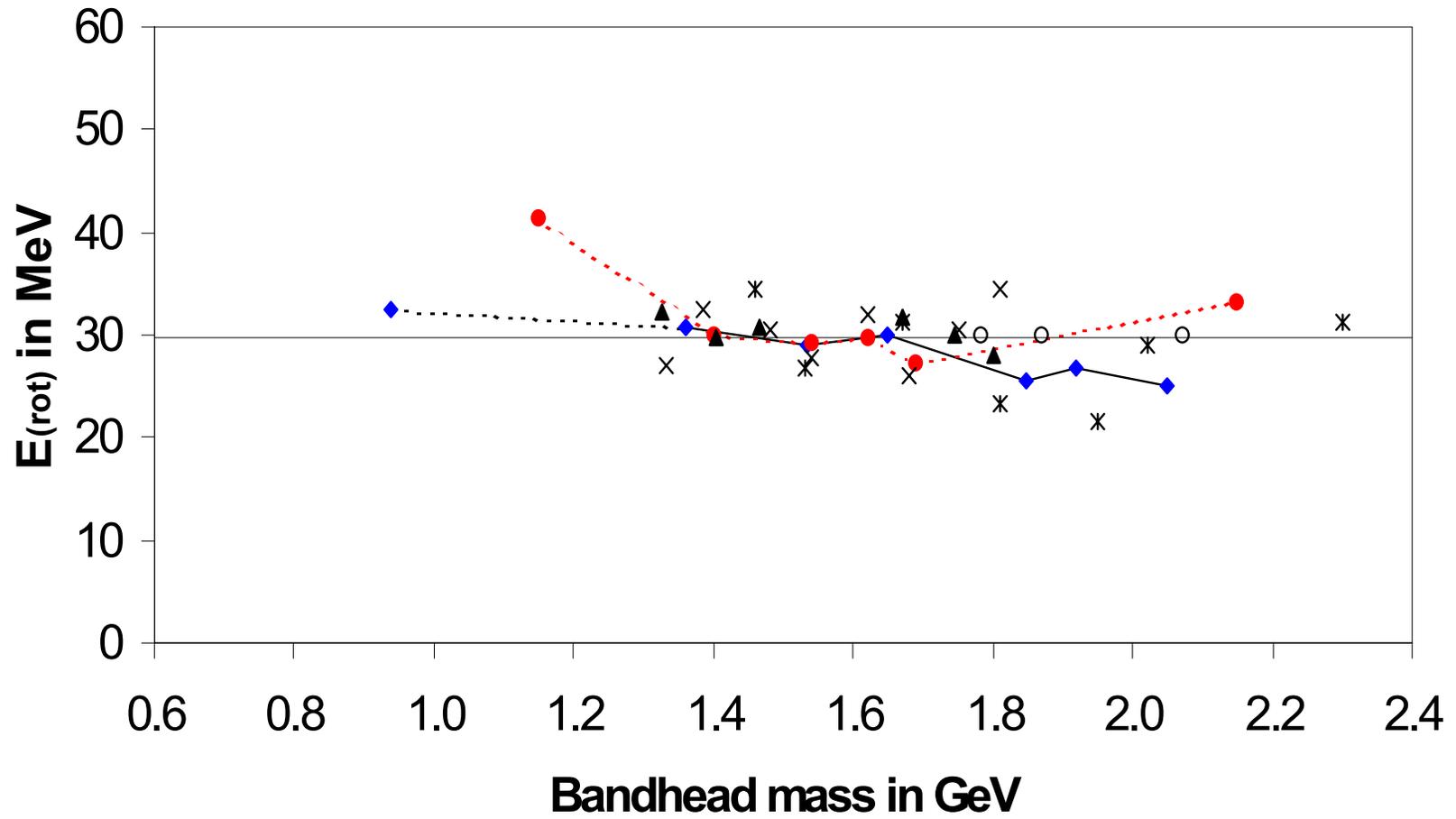

Fig. 6

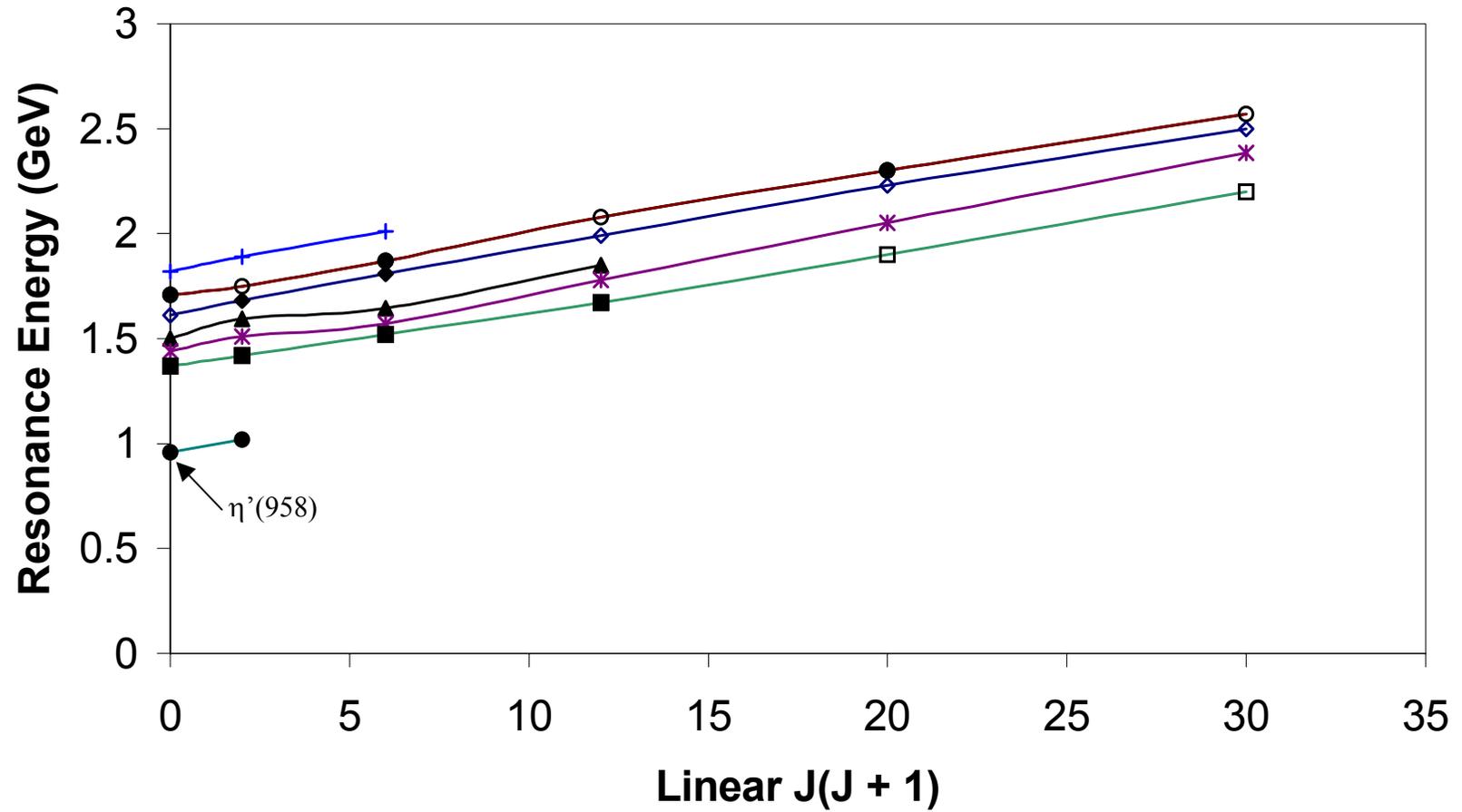

Fig. 7

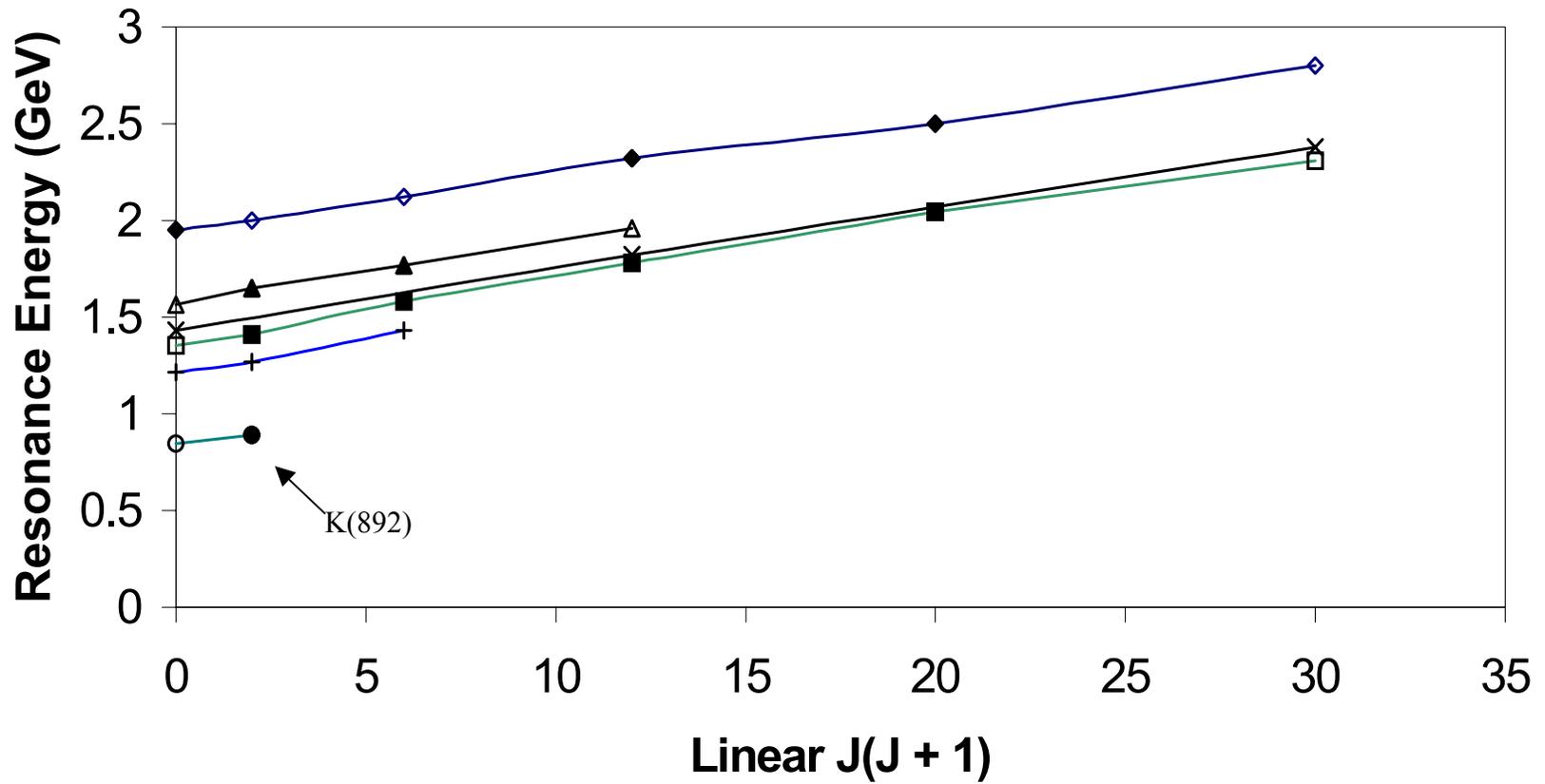

Fig. 8

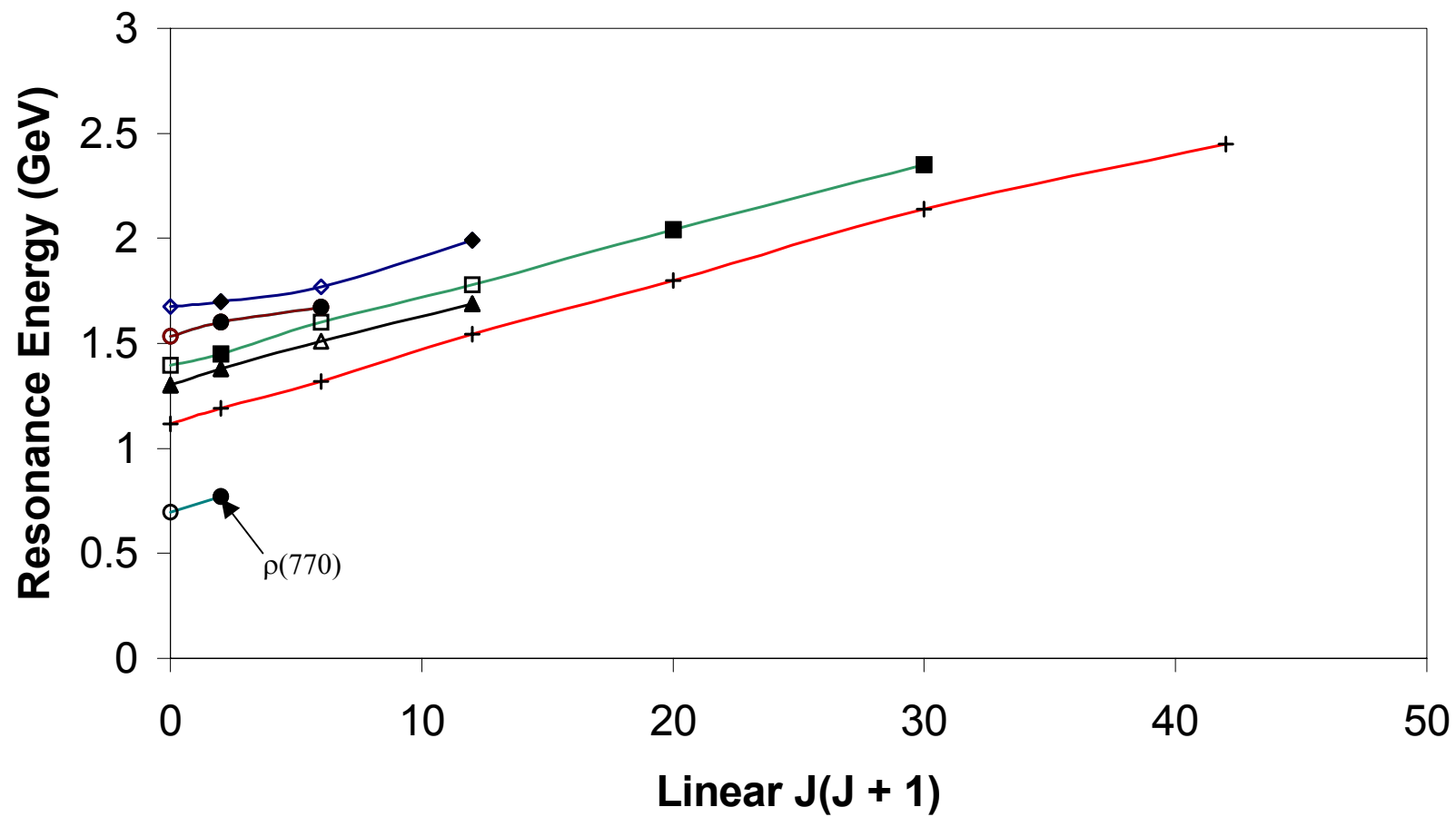

Fig. 9

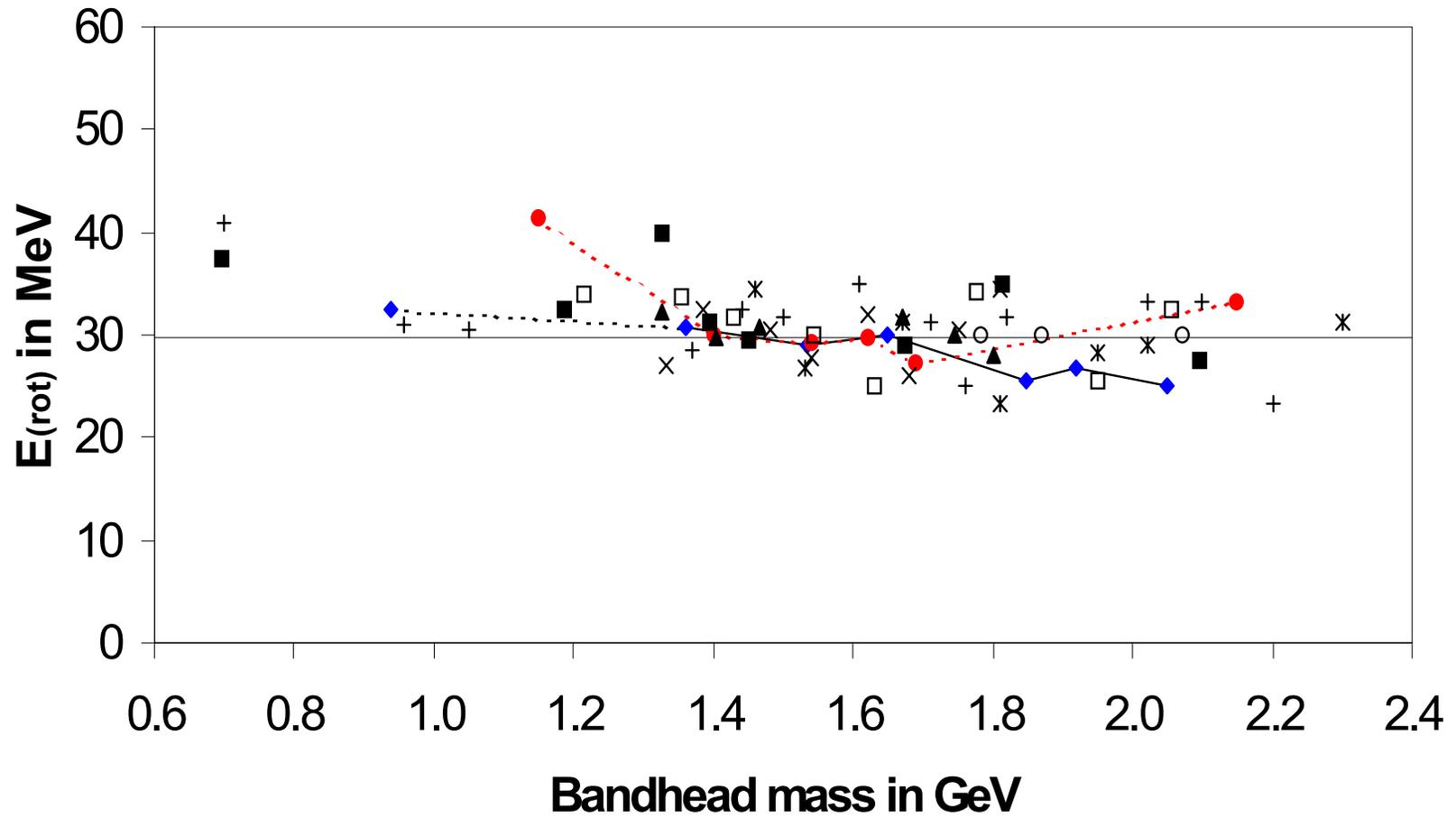

Fig. 10

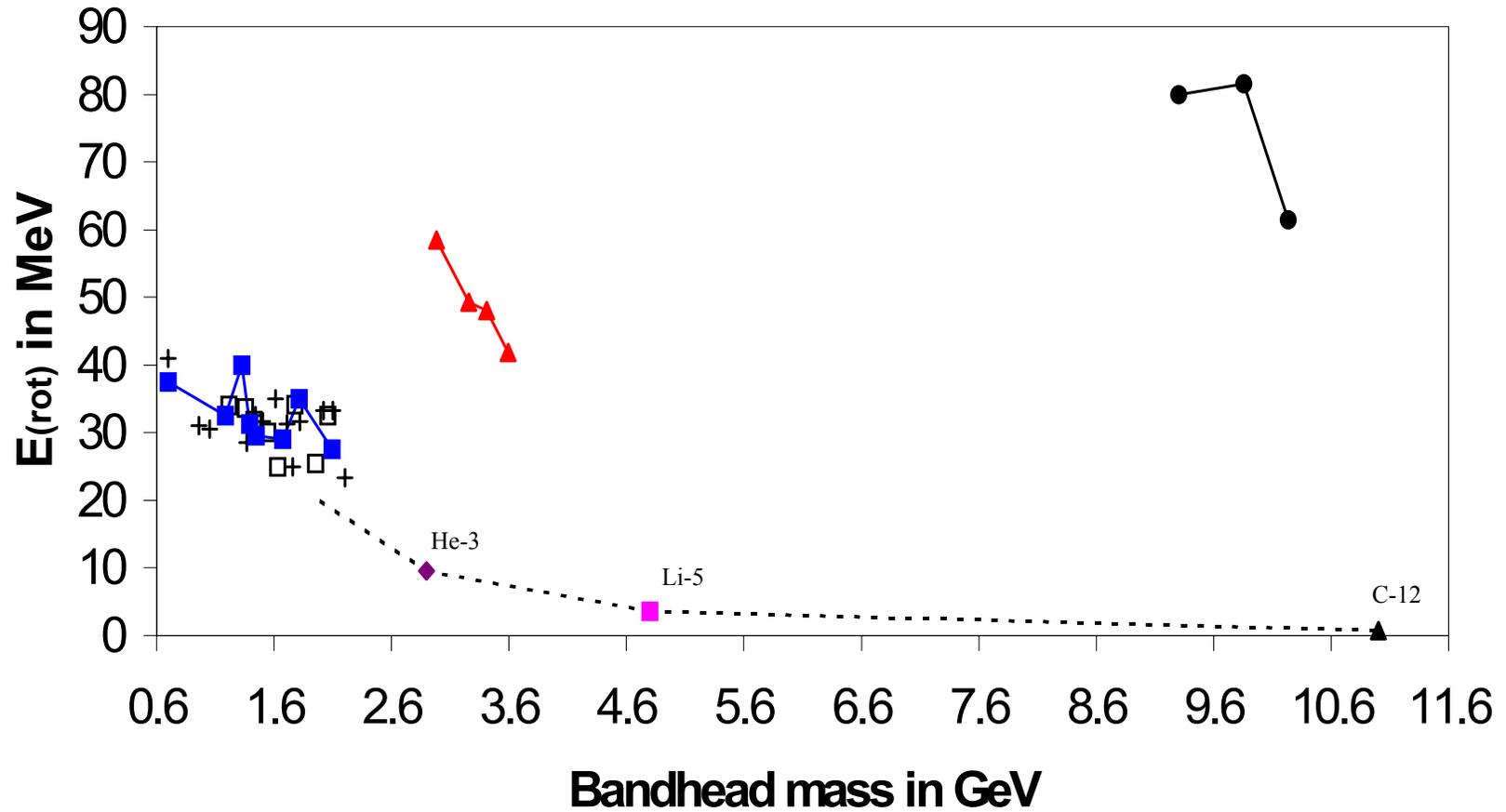

Fig. 11

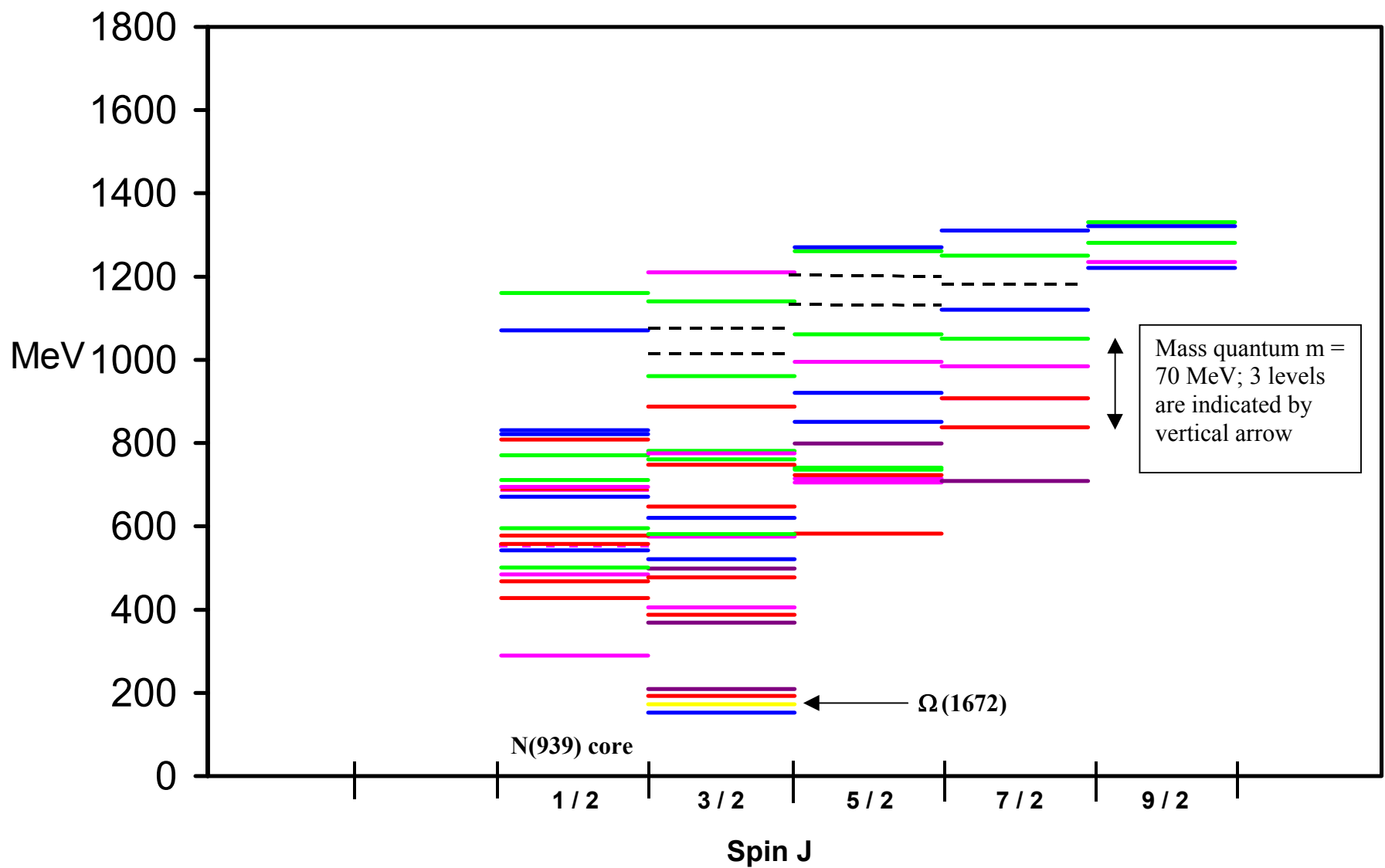

Fig. 12

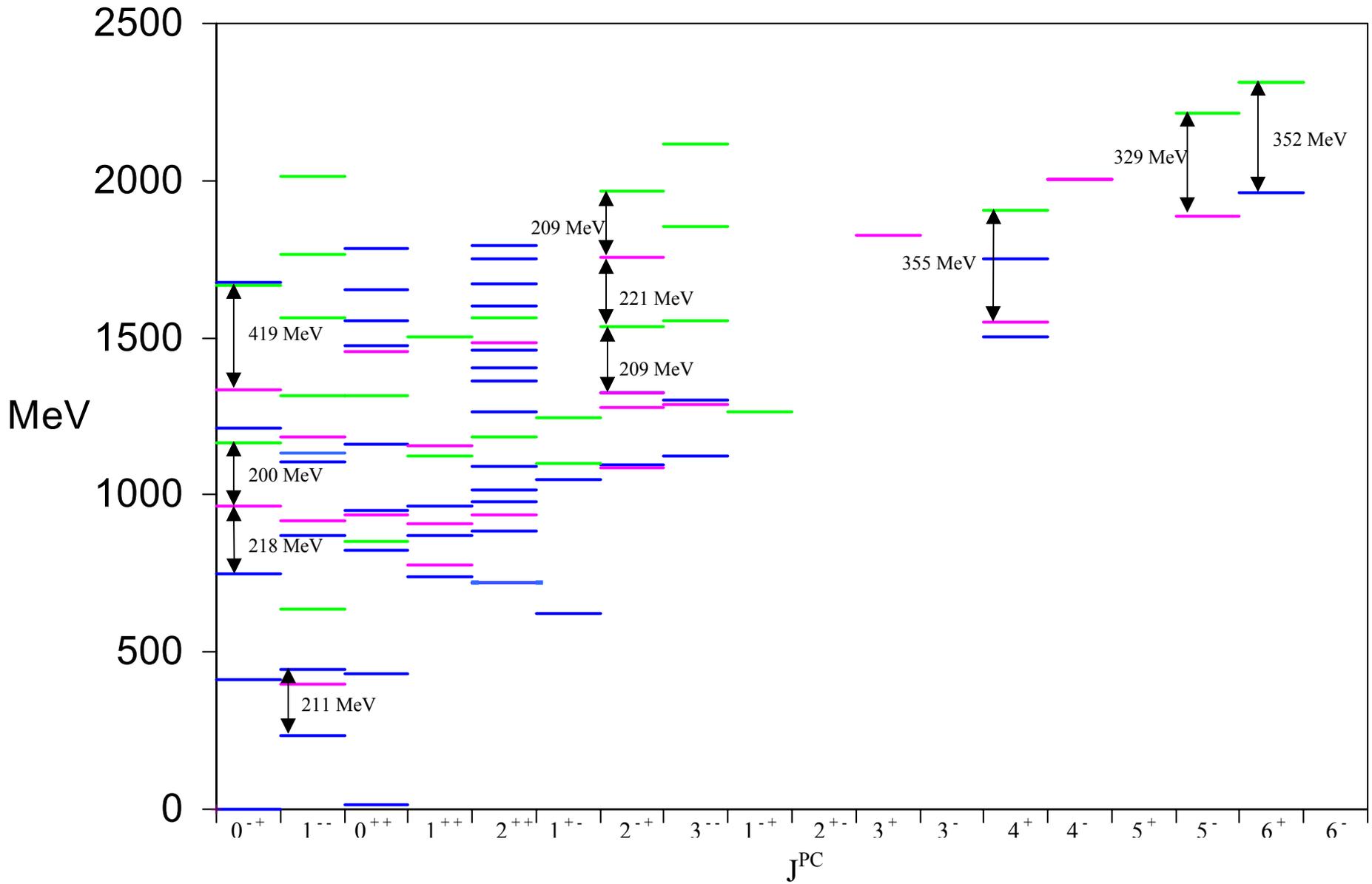

Fig. 13

Table I. Measured values of nucleon masses are listed as a function of angular momentum L. Predicted states are indicated in red. The calculated bandhead energies $E_{calc}$ are from the constituent-quark (CQ) model of Mac Gregor. Rotational energies are calculated between adjacent states and are averaged under the column for $E_{rot}$.

| | | L | 0 | 1 | 2 | 3 | 4 | 5 |
|---|---|---|---|---|---|---|---|---|
| CQ Model | $E_{calc}$ (MeV) | $E_{rot}$ (MeV) Avg. | S | P | D | F | G | H |
| N | 939 | 32.5 | (939) | (1004) | | | | |
| NF | 1149 | | (1149)$S_{11}$ | | | | | |
| NBB | 1219 | | (1219)$S_{11}$ | (1300)$P_{11}$ | (1462)$D_{13}$ | | | |
| NX | 1359 | 30.8 | (1359)$S_{11}$ 40.5 | (1413) (1440)$P_{11}$ 26.8 | (1520)$D_{13}$ 26.7 | (1680)$F_{15}$ 28.8 | (1910)$G_{17}$ 31.0 | (2220)$H_{19}$ |
| NXB | 1499 | 28.9 | (1535)$S_{11}$ 22.5 | (1580)$P_{13}$ 30.0 | (1700)$D_{13}$ 30.0 | (1880)$F_{15}$ 38.75 | (2190)$G_{17}$ 23.0 | (2420)$H_{19}$ |
| NXBB | 1639 | 29.9 | (1650)$S_{11}$ 30.0 | (1710)$P_{11}$ 30.0 | (1830)$D_{13}$ 28.3 | (2000)$F_{15}$ 31.25 | (2250)$G_{19}$ 30.0 | (2550)$H_{19}$ |
| NXXm | 1849 | 25.5 | (1849)$S_{11}$ | (1900)$P_{13}$ | | | | |

| | | | | | |
|---|---|---|---|---|---|
| NXXB | 1919 | 26.8 | (1919)$S_{11}$ | | (2080)$D_{13}$ |
| NXXF | 1989 | | | | |
| NXXFm | 2059 | 25.0 | (2050)<br>(2090)$S_{11}$<br>25.0 | (2100)$P_{11}$<br>25.0 | (2200)$D_{15}$ |

Table II. Measured values of the delta baryon masses are listed as a function of angular momentum L. Predicted states are indicated in red. The calculated bandhead energies $E_{calc}$ are from the constituent-quark (CQ) model of Mac Gregor. Rotational energies are calculated between adjacent states and are averaged under the column for $E_{rot}$.

| | | | L | 0 | 1 | 2 | 3 | 4 | 5 |
|---|---|---|---|---|---|---|---|---|---|
| CQ Model | $E_{calc}$ (MeV) | $E_{rot}$ (MeV) Avg. | | S | P | D | F | G | H |
| Δ | 1079 | | | (1079) | | | | | |
| Δm | 1149 | 41.5 | | (1149)$S_{31}$ | (1232)$P_{33}$ | | | | |
| ΔB | 1219 | | | (1232) | | | | | |
| ΔF | 1289 | | | | | | | | |
| ΔBB | 1359 | | | | | | | | |
| ΔBBm | 1429 | 30.0 | | (1400)$S_{31}$ | | | | (2000)$G_{37}$ | (2300)$H_{39}$ |
| ΔX | 1499 | | | | | | | | |

| | | | | | | | |
|---|---|---|---|---|---|---|---|
| ΔXm | 1569 | 29.3 | (1540)S$_{31}$ 30.0 | (1600)P$_{33}$ 25.0 | (1700)D$_{33}$ 34.2 | (1905)F$_{35}$ 29.4 | (2140)G$_{37}$ 28.0 | (2420)H$_{3,11}$ |
| ΔXB | 1639 | 29.7 | (1620)S$_{31}$ 30.0 | (1680)P$_{33}$ 30.0 | (1800)D$_{35}$ 25.0 | (1950)F$_{37}$ 31.25 | (2200)G$_{37}$ 32.0 | (2520)H$_{39}$ |
| ΔXF | 1709 | 27.2 | (1690)S$_{31}$ 30.0 | (1750)P$_{31}$ 30.0 | (1870)D$_{33}$ 21.7 | (2000)F$_{35}$ | | |
| ΔXBB | 1779 | | | | | (1930)D$_{35}$ | | |
| ΔXBF | 1849 | | | | (1910)P$_{31}$ | | (2400)G$_{39}$ | |
| ΔXX | 1919 | | (1900)S$_{31}$ | | | | | |
| ΔXXm | 1989 | | | | | | | |
| ΔXXB | 2059 | | | | | | (2390)F$_{37}$ | |
| ΔXXF | 2129 | 33.3 | (2150)S$_{31}$ | | (2350)D$_{35}$ | | | |

Table III. Measured values of the lambda baryon masses are listed as a function of angular momentum L. Predicted states are indicated in red. The calculated bandhead energies $E_{calc}$ are from the constituent-quark (CQ) model of Mac Gregor. Rotational energies are calculated between adjacent states and are averaged under the column for $E_{rot}$.

| CQ Model | $E_{calc}$ (MeV) | $E_{rot}$ (MeV) Avg. | L<br>0<br>S | 1<br>P | 2<br>D | 3<br>F | 4<br>G | 5<br>H |
|---|---|---|---|---|---|---|---|---|
| $\Lambda$ | 1116 | | (1116) | | | | | |
| $\Lambda F$ | 1326 | 32.3 | (1326) | (1386)$P_{01}$ | (1520)$D_{03}$ | | | |
| $\Lambda Fm$ | 1396 | 29.8 | (1405)$S_{01}$ | (1460)$P_{01}$ | (1600)$D_{03}$ | | (2000)$G_{07}$ | |
| $\Lambda FB$ | 1466 | 30.7 | (1466) | (1526)$P_{01}$<br>41.0 | (1690)$D_{03}$<br>21.7 | (1820)$F_{05}$<br>35.0 | (2100)$G_{07}$<br>25.0 | (2350)$H_{09}$ |
| $\Lambda X$ | 1536 | | | (1600)$P_{01}$ | | | | |
| $\Lambda Xm$ | 1606 | | | | | | | |
| $\Lambda XB$ | 1676 | 31.7 | (1670)$S_{01}$ | | (1830)$D_{05}$ | (2020)$F_{07}$ | | |
| $\Lambda XF$ | 1746 | 30.0 | | (1810)$P_{01}$ | | (2110)$F_{05}$ | | |

| | | | | (1856) | |
|---|---|---|---|---|---|
| ΛXBB | 1816 | 28.0 | (1800)S$_{01}$ | (1890)P$_{03}$ | (1975)D$_{03}$ |
| ΛXXF | 2166 | | (2145)S$_{01}$ | | (2325)D$_{03}$ |

Table IV. Measured values of the sigma baryon masses are listed as a function of angular momentum L. Predicted states are indicated in red. The calculated bandhead energies $E_{calc}$ are from the constituent-quark (CQ) model of Mac Gregor. Rotational energies are calculated between adjacent states and are averaged under the column for $E_{rot}$.

| | | | L 0 | 1 | 2 | 3 | 4 | 5 |
|---|---|---|---|---|---|---|---|---|
| CQ Model | $E_{calc}$ (MeV) | $E_{rot}$ (MeV) Avg. | S | P | D | F | G | H |
| Σ | 1192 | | (1192) | | | | | |
| Σm | 1261 | | (1262)$S_{11}$ | | | | | |
| ΣB | 1331 | 27 | (1331)$S_{11}$ | (1385)$P_{13}$ | | | | |
| ΣF | 1401 | 32.5 | (1385) | (1445)$P_{13}$ | (1580)$D_{13}$ | | | |
| ΣBB | 1471 | 30.6 | (1480)$S_{11}$ 40.0 | (1560)$P_{11}$ 27.5 | (1670)$D_{13}$ 29.2 | (1845) 25.5 | (2100)$G_{17}$ | |
| ΣBBm | 1541 | 27.8 | (1541) | (1660)$P_{11}$ 25.5 | | (1915)$F_{15}$ 30.0 | | (2455)$H_{19}$ |
| ΣX | 1611 | 31.9 | (1620)$S_{11}$ 35.0 | (1690)$P_{13}$ 21.25 | (1775)$D_{15}$ 39.58 | | (2250)$G_{17}$ | |

| | | | | | | | |
|---|---|---|---|---|---|---|---|
| ΣXm | 1681 | 26.0 | | (1770)$P_{11}$ | | (2030)$F_{17}$ | |
| ΣXB | 1741 | 30.6 | (1750)$S_{11}$ 45.0 | (1840)$P_{13}$ 25.0 | (1940)$D_{13}$ 21.7 | (2070)$F_{15}$ 30.6 | (2620)$H_{1?}$ |
| ΣXF | 1811 | 34.5 | (1811) | (1880)$P_{11}$ | | | |
| ΣXBB | 1881 | | | (1950)$P_{11}$ | | | |
| ΣXFB | 1951 | | (1955)$S_{11}$ (2000)$S_{11}$ | (2020) | | | |
| ΣXFBm | 2021 | | | (2080)$P_{13}$ | | | |

Table V. Measured values of the cascade baryon masses are listed as a function of angular momentum L. Predicted states are indicated in red. The calculated bandhead energies $E_{calc}$ are from the constituent-quark (CQ) model of Mac Gregor. Rotational energies are calculated between adjacent states and are averaged under the column for $E_{rot}$.

| | | | L | 0 | 1 | 2 | 3 | 4 | 5 |
|---|---|---|---|---|---|---|---|---|---|
| CQ Model | $E_{calc}$ (MeV) | $E_{rot}$ (MeV) Avg. | | S | P | D | F | G | H |
| $\Xi$ | 1321 | | | (1321) | | | | | |
| $\Xi m$ | 1391 | | | (1391)$S_{11}$ | | | | | |
| $\Xi B$ | 1461 | 34.5 | | (1461)$S_{11}$ | (1530)$P_{13}$ | | | | |
| $\Xi F$ | 1531 | 26.7 | | (1530) | | (1690)$D_{13}$ | | | |
| $\Xi BB$ | 1601 | | | (1620)$S_{11}$ | | | | | |
| $\Xi FB$ | 1671 | 31.25 | | (1671) | | (1820)$D_{13}$ 35.0 | (2030)$F_{15}$ 27.5 | (2250)$G_{17}$ | |
| $\Xi X$ | 1741 | | | (1741) | | | | | |

| | | | | | | |
|---|---|---|---|---|---|---|
| ΞXm | 1811 | 23.2 | (1811) | (1870)$P_{13}$ | (1950)$D_{13}$ | |
| ΞXB | 1881 | | (1881) | | | |
| ΞXF | 1951 | 28.2 | (1951) | (2025)$P_{11}$ | (2120)$D_{13}$ | |
| ΞXBB | 2021 | 29.0 | (2021) | (2090)$P_{11}$ | | (2370)$F_{15}$ |
| ΞXXB | 2301 | 31.2 | (2301) | | (2500)$D_{13}$ | |

Table VI. Measured values of the omega baryon masses are listed as a function of angular momentum L. Predicted states are indicated in red. The calculated bandhead energies $E_{calc}$ are from the constituent-quark (CQ) model of Mac Gregor. Rotational energies are calculated between adjacent states and are averaged under the column for $E_{rot}$.

| | | | L | 0 | 1 | 2 | 3 | 4 | 5 |
|---|---|---|---|---|---|---|---|---|---|
| CQ Model | $E_{calc}$ (MeV) | $E_{rot}$ (MeV) Avg. | | S | P | D | F | G | H |
| Ω | 1499 | | | (1499) | | | | | |
| Ωm | 1569 | | | (1570)$S_{01}$ | | | | | |
| ΩB | 1639 | | | (1639)$S_{01}$ | (1672)$P_{03}$ | (1868)$D_{03}$ | | | |
| ΩF | 1709 | | | | | | | | |
| ΩBB | 1779 | 30.1 | | (1782)$S_{01}$ | | (1962)$D_{03}$ | | (2384)$G_{07}$ | |
| ΩFB | 1849 | 30.0 | | (1870)$S_{01}$ | | | | (2470)$G_{07}$ | |
| ΩX | 1919 | | | | | | | | |
| ΩXm | 1989 | | | | | | | | |

| | | | | |
|---|---|---|---|---|
| ΩXB | 2059 | 30.0 | (2070)$S_{01}$ | (2250)$D_{03}$ |

Table VII. Measured values of I = 0 meson masses are listed as a function of angular momentum L. Predicted states are indicated in red. The calculated bandhead energies $E_{calc}$ are from the constituent-quark (CQ) model of Mac Gregor. Rotational energies are calculated between adjacent states and are averaged under the column for $E_{rot}$.

| | | | J | 0 | 1 | 2 | 3 | 4 | 5 |
|---|---|---|---|---|---|---|---|---|---|
| CQ Model | $E_{calc}$ (MeV) | $E_{rot}$ (MeV) Avg. | | | | | | | |
| M | 560 | | | η(547) f(560) | | | | | |
| Mm | 630 | | | f(660) | | | | | |
| MB | 700 | 41 | | (700) | ω(782) | | | | |
| MF | 770 | | | | | | | | |
| MBB | 840 | | | f(850) | | | | | |
| MFB | 910 | 31 | | η'(958) | φ(1020) | | | | |
| MX | 980 | | | f(980) | | | | | |

| | | | | | | | |
|---|---|---|---|---|---|---|---|
| MXm | 1050 | 30.5 | | f(1148) | f(1270) | | |
| MXB | 1120 | | (1120) | h(1170) | | | |
| MXF | 1190 | 41.9 | (1190) 47.5 | f(1285) 36.25 | f(1430) | | |
| MXBB | 1260 | 38.7 | η(1295) 62.5 | (1360) ω(1420) 26.3 | f(1525) 27.4 | | f6(2510) |
| MXFB | 1330 | 28.5 | f(1370) 25.0 | f(1420) 36.25 | f(1565) 24.2 | ω(1670) | |
| MXX | 1400 | 32.5 | η(1440) 35.0 | $f_1$(1510) 30.0 | | | f(2050) |
| MXXm | 1470 | 31.7 | $f_0$(1500) 47.5 | h(1595) 12.5 | f(1640) η(1645) 35.0 | φ(1850) | |
| MXXB | 1540 | | η(1540) | ω(1650) | | | |
| MXXF | 1610 | 35 | (1610) | φ(1680) | f(1810) | | |
| MXXBB | 1680 | 31.3 | f(1710) 26.7 | | η(1870) 35.8 | f(2300) | |

| | | | | |
|---|---|---|---|---|
| MXXFB | 1750 | 25.0 | η(1760) | f(1910) f(1950) |
| MX$_3$ | 1820 | 31.7 | η(1820) | f(2010) |
| MX$_3$m | 1890 | | | |
| MX$_3$B | 1960 | | | f(2150) |
| MX$_3$F | 2060 | 33.3 | f(2020) | f$_J$(2220) |
| MX$_3$BB | 2130 | 33.3 | f(2100) | f(2300) |
| MX$_4$ | 2200 | 23.3 | f(2200) η(2225) | f(2340) |
| MX$_4$m | 2270 | | | |
| MX$_4$B | 2340 | | f(2330) | |

Table VIII. Measured values of I = ½ kaon masses are listed as a function of angular momentum L. Predicted states are indicated in red. The calculated bandhead energies $E_{calc}$ are from the constituent-quark (CQ) model of Mac Gregor. Rotational energies are calculated between adjacent states and are averaged under the column for $E_{rot}$.

| | | J | 0 | 1 | 2 | 3 | 4 | 5 |
|---|---|---|---|---|---|---|---|---|
| CQ Model | $E_{calc}$ (MeV) | $E_{rot}$ (MeV) Avg. | | | | | | |
| K | 494 | | (494) (498) | | | | | |
| KF | 704 | | (704) | | | | | |
| KBB | 774 | | (784) | | | | | |
| KBBm | 844 | | | | (892) | | | |
| KX | 914 | | | | | | | |
| KXm | 984 | | | | | | | |
| KXB | 1074 | | | | | | | |

| | | | | | | | |
|---|---|---|---|---|---|---|---|
| KXF | 1144 | | **(1148)** | | | | |
| KXBB | 1214 | 34.0 | (1214) 28.0 | (1270) 40.0 | (1430) | | |
| KXFB | 1284 | | | | | | |
| KXX | 1354 | 33.6 | (1354) 23.0 | (1400) (1410) 45.0 | (1580) 33.3 | (1780) 33.1 | (2045) | (2310) |
| KXXm | 1424 | 31.8 | (1430) 32.5 | | | (1820) 31.1 | | (2380) |
| KXXB | 1494 | | (1460) | | | | |
| KXXF | 1565 | 30.0 | **(1543)** 30.0 | (1650) | (1770) | (1960) | |
| KXXFm | 1635 | 25.0 | (1630) | (1680) | | | |
| KXXFB | 1705 | | | | | | |
| KX$_3$ | 1775 | 34.2 | (1775) | | (1980) | | |

| | | | | | |
|---|---|---|---|---|---|
| KX$_3$m | 1845 | | (1830) | | |
| KX$_3$B | 1915 | 25.5 | (1950) 28.46 | (2320) 22.5 | (2500) |
| KX$_3$Bm | 1985 | | (1985) | | |
| KX$_3$BB | 2055 | 32.5 | (2055) | (2250) | |

Table IX. Measured values of I = 1 meson masses are listed as a function of angular momentum L. Predicted states are indicated in red. The calculated bandhead energies $E_{calc}$ are from the constituent-quark (CQ) model of Mac Gregor. Rotational energies are calculated between adjacent states and are averaged under the column for $E_{rot}$.

| | | J | 0 | 1 | 2 | 3 | 4 | 5 |
|---|---|---|---|---|---|---|---|---|
| CQ Model | $E_{calc}$ (MeV) | $E_{rot}$ (MeV) Avg. | | | | | | |
| M | 135 | | (135) | | | | | |
| MF | 345 | | **(350)** | | | | | |
| MBB | 415 | | | | | | | |
| MBBm | 485 | | | | | | | |
| MX | 555 | | | | | | | |
| MXm | 625 | | | | | | | |
| MXB | 695 | 37.5 | (695) | ρ(770) | | | | |
| MXF | 765 | | | | | | | |
| MXBB | 835 | | | | | | | |
| MXFB | 905 | | | | | | | |
| MXX | 975 | | a(985) | | | | | |
| MXXm | 1045 | | | | | | | |
| MXXB | 1115 | 31.7 | (1115) | | a(1320) | (1545) | (1800) | **a6(2450)** (2140) |
| MXXF | 1185 | 32.5 | (1185) | b(1235) a(1260) 32.5 | (1390) | | | |
| MXXBB | 1255 | | | | | | | |
| MXXFB | 1325 | 40.0 | π(1300) | h(1380) π(1400) | | ρ(1690) | | |
| MX$_3$ | 1395 | 31.9 | | ρ(1450) | | | a(2040) | ρ(2350) |

| | | | | | |
|---|---|---|---|---|---|
| MX$_3$m | 1465 | | a(1450) | | |
| MX$_3$B | 1535 | ? | (1535) | π(1600) | π(1670) |
| MX$_3$F | 1605 | ? | (1605) | a(1640) | a(1700) |
| MX$_3$BB | 1675 | 29.0 | (1675) | ρ(1700) | ρ(1990) |
| MX$_3$FB | 1745 | | | | |
| MX$_4$ | 1815 | | π(1800) | | |
| MX$_4$m | 1885 | 35.0 | | ρ(1900) | ρ(2250) |
| MX$_4$B | 1955 | | | | |
| MX$_4$F | 2025 | | | | |
| MX$_4$BB | 2095 | 27.5 | (2095) | ρ(2150) | |

Table X. Measured values of charmed meson masses are listed as a function of angular momentum L. The calculated bandhead energies $E_{calc}$ are from the constituent-quark (CQ) model of Mac Gregor. Rotational energies are calculated between adjacent states and are averaged under the column for $E_{rot}$.

| | | | J | 0 | 1 | 2 | 3 | 4 | 5 |
|---|---|---|---|---|---|---|---|---|---|
| CQ Model | | $E_{calc}$ (MeV) | $E_{rot}$ (MeV) Avg. | | | | | | |
| M | | 2980 | 58.5 | $\eta_C$(2980) | J/Ψ(3097) | | | | |
| Mm | | 3050 | | | | | | | |
| MB | | 3120 | | | | | | | |
| MF | | 3190 | | | | | | | |
| MBB | | 3260 | 49.3 | (3260) | | | χ(3556) | | |
| MFB | | 3330 | | | | | | | |
| MX | | 3400 | 48.0 | χ(3415) | χ(3511) | | | | |
| MXm | | 3470 | | | | | | | |

| MXB | 3540 | 41.8 | $\eta_C(3594)$ 46.0 | $\Psi(3686)$ 37.5 | $\Psi(3836)$ |

Table XI. Measured values of bottom meson masses are listed as a function of angular momentum L. The calculated bandhead energies $E_{calc}$ are from the constituent-quark (CQ) model of Mac Gregor. Rotational energies are calculated between adjacent states and are averaged under the column for $E_{rot}$.

| | | | J | 0 | 1 | 2 | 3 | 4 | 5 |
|---|---|---|---|---|---|---|---|---|---|
| CQ Model | $E_{calc}$ (MeV) | $E_{rot}$ (MeV) Avg. | | | | | | | |
| M | 9300 | 80.0 | | $\eta_b$(9300) | Y(9460) | | | | |
| MBB | 9580 | | | (9580) | | $\chi_{b2}$(9913) | | | |
| MFB | 9650 | | | | | | | | |
| MX | 9720 | | | (9720) | $\chi_{b1}$(9893) | | | | |
| MXm | 9790 | | | | | | | | |
| MXB | 9860 | 81.5 | | $\chi_{b0}$(9860) | Y(10023) | | | | |
| MXF | 9930 | | | (9930) | | $\chi_{b2}$(10269) | | | |
| MXFm | 10000 | | | | | | | | |

| | | | | |
|---|---|---|---|---|
| MXFB | 10070 | | (10070) | $\chi_{b1}(10255)$ |
| MXX | 10140 | | (10140) | $\chi_{b2}(10479)$ |
| MXXm | 10210 | 61.5 | $\chi_{b0}(10232)$ | Y(10355) |
| MXXB | 10280 | | (10280) | $\chi_{b1}(10465)$ |
| MXXBm | 10350 | | (10350) | |
| MXXBB | 10420 | | $\chi_{b0}(10420)$ | Y(10580) |